\documentstyle[epsf]{mn}

\title{Polarized QPOs from the INTEGRAL polar IGRJ14536-5522 (=Swift
  J1453.4-5524). }

\author[Stephen B. Potter et al.]  {Stephen B. Potter$^{1}$, David
  A. H.  Buckley$^{1}$, Darragh O'Donoghue$^{1}$, \and Encarni
  Romero--Colmenero$^{1}$, James O'Connor$^{1}$, Piet Fourie$^{1}$,
  Geoff Evans$^{1}$, \and Craig Sass$^{1}$, Lisa Crause$^{1}$, Martin
  Still$^{4,5}$, O. W. Butters$^{2,6}$, A.J. Norton$^{2}$ \and and Koji
  Mukai$^{3}$ \\
  $^{1}$South African Astronomical Observatory, PO Box 9,
  Observatory 7935, Cape Town, South Africa \\
  $^{2}$Department of Physics and Astronomy, The Open University, Walton Hall,
  Milton Keynes, MK7 6AA, UK. \\
  $^{3}$CRESST and X-ray Astrophysics Laboratory NASA/GSFC, Greenbelt, MD
  20771, USA; Department of Physics, \\ University of Maryland,  Baltimore
  county,
  1000 Hilltop Circle, Baltimore, MD 21250, USA; \\
  $^{4}$Mullard Space Science Laboratory, University College London, Holmbury
  St
  Mary, Dorking, Surrey RH5 6NT\\
  $^{5}$NASA Ames Research Center, Moffett Field, CA 94035, USA \\
  $^{6}$Department of Physics and Astronomy, University of Leicester, Leicester, LE1 7RH, UK \\
}

\newcommand{\hbeta}{H$\beta\ $}
\newcommand{\hgamma}{H$\gamma\ $}

\date{}

\begin{document}

\maketitle

\begin{abstract}
  
We report optical spectroscopy and high speed photometry and
polarimetry of the INTEGRAL source IGRJ14536-5522 (=Swift
J1453.4-5524). The photometry, polarimetry and spectroscopy are
modulated on an orbital period of $3.1564(1)$ hours. Orbital
circularly polarized modulations are seen from $\sim 0$ to $\sim -18$
per cent, unambiguously identifying IGRJ14536-5522 as a polar. The
negative circular polarization is seen over $\sim 95$ per cent of the
orbit, which is consistent (as viewed from Earth) with a single pole
accretor. We estimate some of the system parameters by modeling the
polarimetric observations.

Some of the high speed photometric data show modulations that are
consistent with quasi-periodic oscillations (QPOs) on the order of 5-6
minutes. Furthermore, for the first time, we detect the (5-6) minute
QPOs in the circular polarimetry. We discuss the possible origins of
these QPOs. In addition, we note that the source undergoes frequent
changes between different accretion states.

We also include details of HIPPO, a new high-speed photo-polarimeter used for
some of our observations. This instrument is capable of high-speed,
multi-filtered, simultaneous all-Stokes observations. It is therefore
ideal for investigating rapidly varying astronomical sources such as
magnetic Cataclysmic Variables.

\end{abstract}

 \begin{keywords}
     accretion, accretion discs -- methods: analytical -- techniques:
     polarimetric -- binaries: close -- novae, cataclysmic variables --
     X--rays: stars. 
 \end{keywords}

\section{Introduction}

The standard picture of a cataclysmic variable (CV) is a binary system
consisting of a Roche lobe filling red dwarf (known as the secondary
or the donor star) and an accreting white dwarf (the primary). CVs
have orbital periods of typically a few hours, and mass transfer is
caused by angular momentum loss - see e.g. Warner (1995) for a review
of cataclysmic variables.  Approximately 20\% of the known CVs are
magnetic cataclysmic variables (mCVs), where the white dwarf has a
strong magnetic field (see the catalogue of Ritter \& Kolb
2003). These are further sub-divided into two subtypes, namely
intermediate polars (IPs) and polars, depending on the strength of the
magnetic field of the white dwarf and the degree of synchronism
between the white dwarf spin and the binary orbit- see the reviews
given by Cropper (1990) and Patterson (1994).

In polars, also known as AM Her systems, the white dwarf has a
sufficiently strong magnetic field to lock the system into synchronous
rotation and to prevent completely the formation of an accretion
disc. Instead, the material from the secondary overflowing the Roche
lobe initially falls towards the white dwarf following a ballistic
trajectory until, at some distance from the white dwarf, the magnetic
pressure overwhelms the ram pressure of the ballistic stream, confining
the flow until it eventually reaches the surface where it forms a hot
shocked region. 

It has long been known that CVs are a significant source of soft ($<$
2 keV) and medium energy (2-10 keV) X-rays (e.g. Patterson et
al. 1984). Recent surveys with INTEGRAL show that CVs are also notable
sources of hard ($>$ 20 keV) X-rays. A large fraction of these are
made up of magnetic CVs and in particular IPs (see Barlow et al. 2006
and Revnivtsev et al 2008). Interestingly, out of the five
asynchronous polars known, two are INTEGRAL sources.

IGRJ14536-5522 (=Swift J453.4-5524) was discovered as a hard X-ray
source by INTEGRAL (Kuiper, Keek, Hermsen, Jonker \& Steeghs 2006) and
by Swift/BAT (Mukai et al. 2006). A pointed Swift/XRT observation led
to the identification with a ROSAT all-sky survey (RASS) source 1 RXS
J145341.1-552146, and hence to its optical identification (Masetti et
al. 2006). Revnivtsev et al. (2008) classify it as an IP.

Follow-up spectroscopy and photometry with SALT (Southern African
Large Telescope) and with the SAAO (South African Astronomical
Observatory) 1.9m telescope showed that this object belongs to a rare
subtype of magnetic CV: a hard X-ray bright polar or a soft
intermediate polar (Mukai, Markwardt, Tueller, Buckley, Potter, Still
et al. 2006). Further observations were clearly needed.

Here we report on our optical photo-polarimetric and spectroscopic
observations of IGRJ14536-5522. The paper is structured as follows: In
Sect. 2 we describe the overall design of a new high speed
photo-polarimeter used for some of our observations. Sect. 3 gives an
account of all our observations, followed by our analysis of the
spectroscopic and photo-polarimetric observations in Sects. 4 and 5
respectively. We finish with a discussion and summary in Sect. 6.

\section{The High speed Photo-POlarimeter (HIPPO)}

SAAO's HIPPO was designed and built in order to replace its highly
successful but aging single channel equivalent, namely the UCT
(University of Cape Town) photo-polarimeter (Cropper 1985).  Its
purpose is to obtain simultaneous all-Stokes parameters,
multi-filtered observations of unresolved astronomical sources. In
addition, it is capable of high speed photo-polarimetry in order to
permit investigations of rapidly varying polarized astronomical
sources. Of particular interest are magnetic Cataclysmic Variables
(mCVs). This is the first refereed publication of the instrument and
therefore we describe the overall instrument design, data acquisition
and reduction here.

\subsection{The optical design}

\begin{figure*}
\epsfxsize=16.5cm
\epsffile{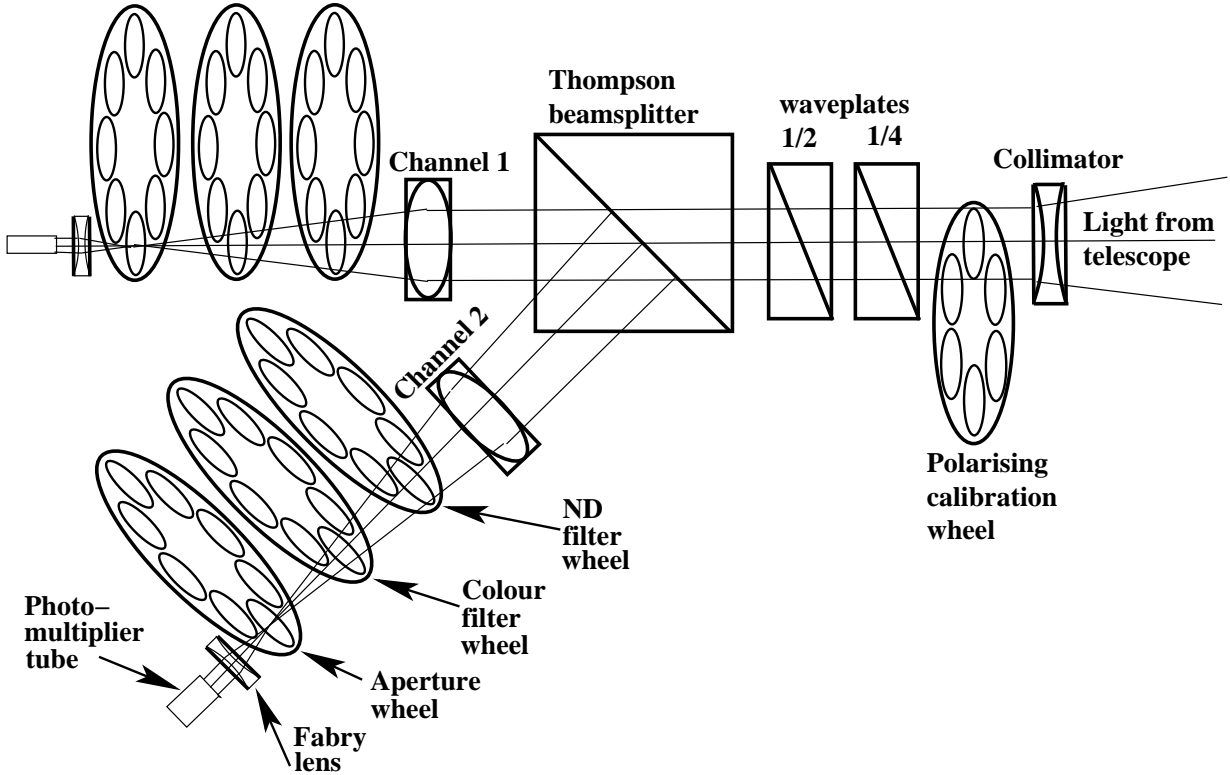} 
\caption{Optical layout of the polarimeter. Channel 1 is a copy of channel 2.}
\label{opt_lay}
\end{figure*}

Fig.~\ref{opt_lay} shows a schematic diagram of the optical layout of the
polarimeter.  Light from the telescope first encounters a field lens that
produces a collimated beam. Within the collimated beam is placed a polarizing
calibration filter wheel followed by super-achromatic 1/4 and 1/2 wave-plates.
The polarizing calibration filter wheel consists of 2 linear polaroids, 1
circular polaroid (a linear followed by a 1/4 wave plastic retarder to produce a
circularly polarized beam), a Lyot depolarizer and an open position. These
filters are used for calibration and efficiency measurements of the instrument
and/or the telescope. The 1/4 and 1/2 wave-plates are also placed in the
collimated beam in order to minimise any lateral modulation of the pupil image
as the wave-plates are rotated. A Thompson beam-splitter then produces the
ordinary and extraordinary beams. All of the above polarizing optics are placed
before any filters or apertures, to avoid problems caused by metallic
apertures or filters with residual stress birefringence.

Each beam has its own neutral density, colour filter and aperture wheels.
The beams are focused at the aperture wheels by lenses at the top of each
channel. Fabry lenses re-image the pupil onto two photo-multiplier tubes. There
is also an eye-piece (used for initial alignment) and a dark slide on each
channel.

The wave-plates are contra-rotated at 10Hz and therefore modulate the
ordinary and extraordinary beams. The modulation is sufficiently rapid
that errors which arise as a result of variable atmospheric conditions
or telescope guiding modulations are much reduced. In addition,
modulations that occur as a result of wedge shaped rotating elements,
dirt on the rotating components or dichroism from refraction at the
element surfaces appear mostly at harmonics that do not affect the
measurement of polarization. The remaining sources of error are photon
statistics, which can be minimised by collecting a larger number of
photons, and instrument/telescope polarization, which can be measured
by observing polarized and un-polarized standard stars in combination
with the calibration filters. This recipe for measuring
polarimetry is based on the work of Serkowski (1974). Measurements of
all the Stokes parameters are made simultaneously from the modulated
beams. As both beams are modulated, each provides an independent
measurement of the polarization. Therefore, simultaneous two filter
observations are possible. Linear- and circular-only modes are also
possible by rotating only the 1/2 or 1/4-wave-plates respectively.

Serkowski (1974) provides the formalism for calculating the Stokes
parameters from O and E beams that are modulated as a result of
passing through two retarders in series. In our case, for constantly
contra-rotating 1/4 and 1/2 wave-plates in series, the modulated
intensities are given by


\begin{eqnarray*}
I_{O}^{'} = {1\over 2} \Big( I+{1\over
  2}Q\big[cos8\Psi_{8}+cos4\Psi_{4}\big]+{1\over
  2}U\big[sin8\Psi_{8}-sin4\Psi_{4}\big] \\
  -V\big[sin6\Psi_{6}\big]  \Big)
\end{eqnarray*}

\begin{eqnarray*}
I_{E}^{'} = &&{1\over 2} \Big( I+{1\over
  2}Q\big[-cos8\Psi_{8}-cos4\Psi_{4}\big]+ \\
&&{1\over 2}U\big[-sin8\Psi_{8}+sin4\Psi_{4}\big] 
+V\big[sin6\Psi_{6}\big]  \Big)
\end{eqnarray*}

for the O and E beams respectively. Where
\begin{eqnarray*}
\Psi_{8}&=&8\Psi-4C^{({1\over2})}+4C^{({1\over4})} \\
\Psi_{6}&=&6\Psi-4C^{({1\over2})}+2C^{({1\over4})} \\
\Psi_{4}&=&4\Psi-4C^{({1\over2})} 
\end{eqnarray*}
and $\Psi$ is the angle between the fast axis of the two wave-plates
and referenced to the 1/4 wave-plate fast axis. $C^{({1\over2})}$ and
$C^{({1\over4})}$ are the zero point constant offsets for the 1/2 and
1/4 wave-plates respectively, and I, Q, U and V are the Stokes
parameters. From these equations one can see that the linear component
of the polarization is modulated equally at the 4th and 8th harmonics
of the rotation frequency. The circular component is modulated at the
6th harmonic. The linear polarization is measured by adding the
amplitudes of the 4th and 8th harmonics and, similarly by measuring
the 6th harmonic for the circular polarization. A least squares
algorithm is used to obtain the amplitudes and phases of the
harmonics. Correction factors are applied to each harmonic in order to
compensate for the fact that the modulated signal is made up of a
finite number of bins (100). Efficiency factors, to compensate for the
slight wavelength dependence in retardance of the wave-plates and
instrumental polarization, are measured by observing polarized and
unpolarized standard stars.

\subsection{Data acquisition and reductions}

The control and data acquisition software is written in C and is
hosted by an industrial PC. The photometer counts (X2), minute and
second time pulses and 1/2+1/4 waveplate pulses are handled by real
time C code in order to ensure correct and absolute timely recording
of the data. The real time code is driven by a 1 milli-second time
interval interrupt driven by a 1KHz signal from the time service
provided by the Observatory. At every interrupt, the status of the
waveplate pulses, time pulses and the photometer buffers are
recorded. These data are then sent to the user C code, where
on-the-fly data reductions are performed.  The 1 milli-second data
stream is also saved to disk for later off-line data reduction. With a
data rate of 1KHz, the 10Hz rotating wave plates are sampled 100 times
per revolution. Therefore every 0.1 seconds a polarization measurement
is made. Off-line data reductions permit binning of the data to any
integer multiple of 0.1 seconds.

\section{Observations}

 Table ~\ref{tab:observations} shows a log of all the observations of
 IGRJ14536-5522.

\begin{table*}
\begin{center}
\caption{Table of observations. Cass Spect is the Cassegrain
  spectrograph, 1.9m and HIPPO are the 1.9m telescope and the HIgh
  speed Photo-POlarimeter respectively of the South African
  Astronomical Observatory. OG570 and BG390 are broad-band red and
  blue filters respectively. SALTICAM is the CCD camera of the
  Southern African Large Telescope. UCTCCD is the University of Cape
  Town CCD camera. RSS is the Robert Stobie Spectrograph. Observed
  bright and faint states are indicated by $^{h}$ and $^{l}$ in the
  last column. $^{M}$ denotes observations originally published in
  Mukai et al. 2006.  {\label{tab:observations}}} \vspace{0.2cm}
\centerline{
\begin{tabular}{|r|c|c|c|c|c|c|r|r|} \hline
Date  & Telescope & Instrument & Spectral range/ & Resolution  & No. of & Integration & Dataset length \\
      &           &            &     filter &(\AA)& spectra & times(s) & (orbits) \\ \hline
14/15   Sep 2006 & 1.9m & UCT CCD & Clear &-& - & 60 & $\sim$ 0.85$^{hM}$ \\
15/16   Sep 2006 & 1.9m & UCT CCD & Clear &-& - & 60 & $\sim$ 0.37$^{hM}$ \\
17/18   Sep 2006 & 1.9m & UCT CCD & Clear &-& - & 60 & $\sim$ 0.72$^{hM}$ \\
18/19   Sep 2006 & 1.9m & UCT CCD & Clear &-& - & 60 & $\sim$ 0.82$^{hM}$ \\
14-9 May/Jun 2006 & SALT & RSS      & 5930-7212\AA   &1& 31  &600& $\sim$ 0.9$^{lM}$\\
6/7   Jul 2007 & 1.9m & Cass Spect & 4000-5000\AA &1& 42 & 600 & $\sim$ 2.1$^{h}$ \\
8/9   Jul 2007 & 1.9m & Cass Spect & 4000-5000\AA &1& 42 & 600 & $\sim$ 2.1$^{h}$ \\
9/10  Jul 2007 & 1.9m & Cass Spect & 4000-5000\AA &1& 39 & 600 & $\sim$ 2.0$^{h}$ \\
10/11 Jul 2007 & 1.9m & Cass Spect & 3750-7800\AA &4& 42 & 500 & $\sim$ 2.0$^{h}$ \\
25/26 Jul 2007 & 1.9m & Cass Spect & 4000-8000\AA &4& 9  & 1200 & $\sim$ 1.0$^{l}$ \\
27/28 Feb 2008 & 1.9m & HIPPO      & Unfiltered   &-& -  &1ms,0.1& $\sim$ 1.2$^{h}$ \\
28/29 Feb 2008 & 1.9m & HIPPO      & OG570, BG39   &-& -  &1ms,0.1& $\sim$ 1.2$^{h}$\\
29/1 Feb 2008 & SALT & SALTICAM      & Clear   &-& -  &0.1& $\sim$ 0.35$^{h}$\\
1/2 Mar 2008 & SALT & SALTICAM      & Clear   &-& -  &0.1& $\sim$ 0.15$^{h}$\\
1/2 Mar 2008P & 1.9m & HIPPO      & B,I   &-& -  &1ms& $\sim$ 1.2$^{h}$ \\
2/3 Mar 2008 & SALT & SALTICAM      & Clear   &-& -  &0.1& $\sim$ 0.17$^{h}$\\
2/3 Mar 2008 & 1.9m & HIPPO      & B,R   &-& -  &1ms& $\sim$ 1.2$^{h}$ \\
3/4 Apr 2008 & 1.9m & HIPPO      & B,I   &-& -  &1ms, 0.1s& $\sim$ 1.2$^{h}$ \\
5/6 Apr 2008 & 1.9m & HIPPO      & B,V,R,I   &-& -  &1ms, 0.1s& $\sim$ 1.5$^{h}$ \\
7/8 Apr 2008 & 1.9m & HIPPO      & Clear,B,I   &-& -  &1ms, 0.1s& $\sim$ 1.2$^{h}$ \\
7/8 May 2008 & SALT & SALTICAM      & U   &-& -  &0.1& $\sim$ 0.13$^{h}$\\
7/8 May 2008 & SALT & SALTICAM      & B   &-& -  &0.1& $\sim$ 0.26$^{h}$\\
9/10 May 2008 & SALT & SALTICAM      & V   &-& -  &0.1& $\sim$ 0.23$^{h}$\\
13/14 May 2008 & SALT & SALTICAM      & R   &-& -  &0.1& $\sim$ 0.3$^{h}$\\
17/18 May 2008 & SALT & SALTICAM      & B   &-& -  &0.1& $\sim$ 0.26$^{h}$\\
18/19 May 2008 & SALT & SALTICAM      & V   &-& -  &0.1& $\sim$ 0.1$^{h}$\\
\hline
\end{tabular}
}
\end{center}
\end{table*}

\subsection{Optical Spectroscopy}

Spectroscopic observations of IGRJ14536-5522 were made during July
2007, on the 1.9-m telescope located on the Sutherland site of the
South African Astronomical Observatory (SAAO), using the Cassegrain
spectrograph with the SITe1 CCD (1752 $\times$ 266 $\times$ 15 $\mu$m
pixels). The higher resolution grating was chosen in order to cover
the \hbeta, HeII and \hgamma emission lines. The lower resolution
grating was chosen in order to cover a broader optical wavelength
range. Flat field spectra were obtained at the beginning and/or end of
each night and wavelength calibration was provided by observing a CuAr
lamp approximately every 20-25 minutes.  Spectrophotometric flux
standards were also observed, allowing flux calibration. Data
reductions made use of the standard tools available through IRAF.

SALT RSS spectra were obtained when the system was in a lower state
between May 14 and Jun 9 2006. These observations were presented in
Mukai et al. (2006) and have been listed here for completeness and for
comparison.

\subsection{Optical Polarimetry}

IGRJ14536-5522 was observed polarimetrically on the SAAO 1.9m
telescope during the commissioning week of the HIPPO in February and
March 2008 and then later in April 2008. The HIPPO was operated in its
simultaneous linear and circular polarimetry and photometry mode
(All-Stokes). White light observations (3500-9000\AA) were defined by
the response of the two RCA31034A GaAs photomultiplier tubes, whilst
for others a broad blue band (3500-5500\AA) BG39 filter, a broad red
band (5700-9000\AA) OG570 filter, or B,V,R,I filters were used.  

Several polarized (HD80558, HD111579, HD111613, HD147084, HD126593,
HD147084, HD160529, HD298383, HD110984) and non-polarized (HD90156,
HD100623) standard stars (Hsu \& Breger 1982 and Bastien et al. 1988)
were observed in order to calculate the position angle offsets,
instrumental polarization and efficiency factors. Background sky
polarization measurements were also taken at frequent intervals during
the observations.  Data reduction then proceeded as outlined in the
previous sections.

\subsection{Optical Photometry}

IGRJ14536-5522 was observed photometrically with the UCTCCD in 2006,
as part of the polarimetric observations in 2008 and with SALTICAM in
2008. The observations are not absolutely photometrically calibrated.


\section{Spectroscopic analysis}

\subsection{The mean spectrum}

\begin{figure}
\epsfxsize=8cm
\epsffile{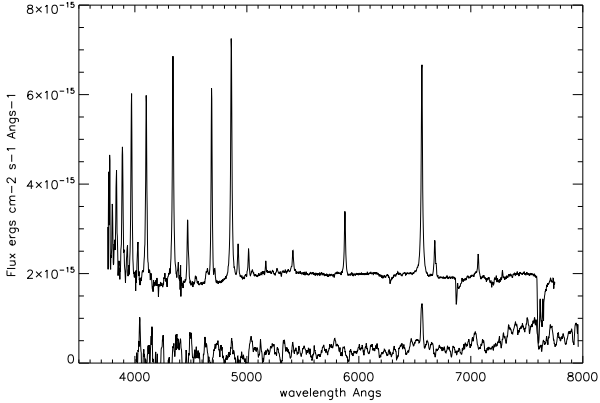} 
\caption{Average of the low resolution spectra. Upper and lower
  average spectra are made from the observations taken on 10/11
  (during a higher state) and the 25/26 (during a lower state) July
  2007, respectively. The lower state data has been magnified by 10.}
\label{average_spectG7}
\end{figure}



\begin{figure}
\epsfxsize=8cm
\epsffile{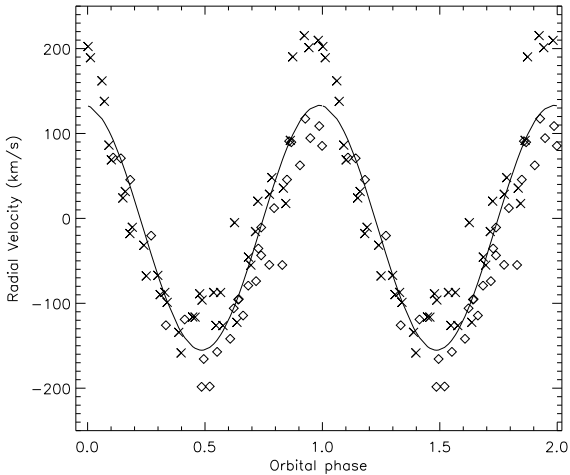} 
\caption{ The H$\alpha$ radial velocity curve of IGRJ14536-5522 folded on the
  3.1564 hr period with arbitrary phasing. Crosses and diamonds are
  the higher state 2007 and the lower state 2006 observations
  respectively. }
\label{chi_stat}
\end{figure}

Fig.~\ref{average_spectG7} shows the average of the spectra taken when
the system was in a higher state (upper spectrum) and when it was
observed to be in a lower state two weeks later during July of 2007
(lower spectrum). These observations were made with the SAAO 1.9m
telescope. The higher state spectrum is very typical for a polar (see
e.g. QS Tel: Rosen et al. 1996, HU Aqr: Schwope, Mantel \& Horne,
1997), i.e. it exhibits Balmer emission lines (H$\alpha$ - H$11$) as
well as neutral helium (He I $\lambda\lambda$ 4026, 4387, 4471, 4713,
4922, 5016, 5048, 5876, 6678, 7065\AA), ionised helium (He II
$\lambda\lambda$ 4542, 4686, 5411). CaII (3933\AA), blended CIII, OII,
NIII (4640\AA) and FeI (5172\AA). The phase-resolved low-resolution
spectra clearly display bright and faint phases (not shown).




\subsection{A spectroscopic ephemeris for the secondary.}

We measured the radial velocity of the H$\alpha$ emission line in our
July 2007 observations using a single Gaussian convolution
method. These results were then combined with the same measurements
made by Mukai et al (2006) of their June 2006 observations. A
$\chi^{2}$ minimization technique was then used in order to search for
any periods. A period of $3.1564(1)$ hours was detected which is
consistent with that found by Mukai et al (2006) and recognized as the
orbital period. The period error arises as a result of not being able
to distinguish between aliases.

Fig.~\ref{chi_stat} shows the H$\alpha$ radial velocity curve folded
on the 3.1564 hr period with arbitrary phasing. We measure a gamma
velocity of $-7.3(1)$ km/s and a semi-amplitude of $144.9(2)$ km/s.
The numbers in brackets are the one sigma standard deviations of a
sine fit to the radial velocities.  The higher and lower state data
are distinguished by the crosses and diamonds respectively. Note that
the higher state data appears to have a generally higher velocity than
the lower state data. We attribute this to different emission regions
contributing to the line emission when in different states: the
irradiated face of the secondary star dominates the emission line
profile during lower states, but the emission lines become
multi-component during higher states (see e.g. UW Pic:
Romero-Colmenero et al. 2003). Hence the multi-component nature of the
emission lines have probably biased our gamma velocity measurement
due to asymmetric line profiles.

In order to derive a zero point for the ephemeris we analysed our
higher resolution spectroscopic data.  Fig.~\ref{trl_dop} shows these
data, taken in July 2007, phased and folded on the above detected
period and centered on the HeII 4686\AA~line. These observations show
the multi-component nature more clearly (described in more detail
below). In particular, a bright narrow component can be seen
(indicated by the dashed curve), which is commonly associated with the
irradiated face of the secondary star. We fitted multi(3)-Gaussian
profiles to the trailed spectra and calculated the radial velocities
of the 3 visible components as a function of phase. A best sinusoidal
fit to the narrow component was used in order to derive the time of
blue-red crossing and hence the time of inferior conjunction of the
secondary star. Accordingly, this in turn is used to give the epoch to
our ephemeris:

$$ T(HJD) = 2454290.14723(8) + 0.131517(4)E $$

The number enclosed in brackets (on the epoch) is the formal one sigma
error measurement. However we note that the true superior conjunction
of the secondary may be offset by a value that is larger than the
quoted error if the irradiated face of the secondary is not symmetric
about the line of centers of the two stars.

We measure a gamma velocity and semi-amplitude of $-26.0(3)$ km/s
and $80.8(4)$ km/s respectively for the narrow component. The
numbers in brackets are the one sigma standard deviations of a sine
fit to the radial velocities. Henceforth, all our observations are
phased on the above ephemeris.

\subsection{The trailed spectra and Doppler tomograms}

\begin{figure*}
\epsfxsize=16cm
\epsffile{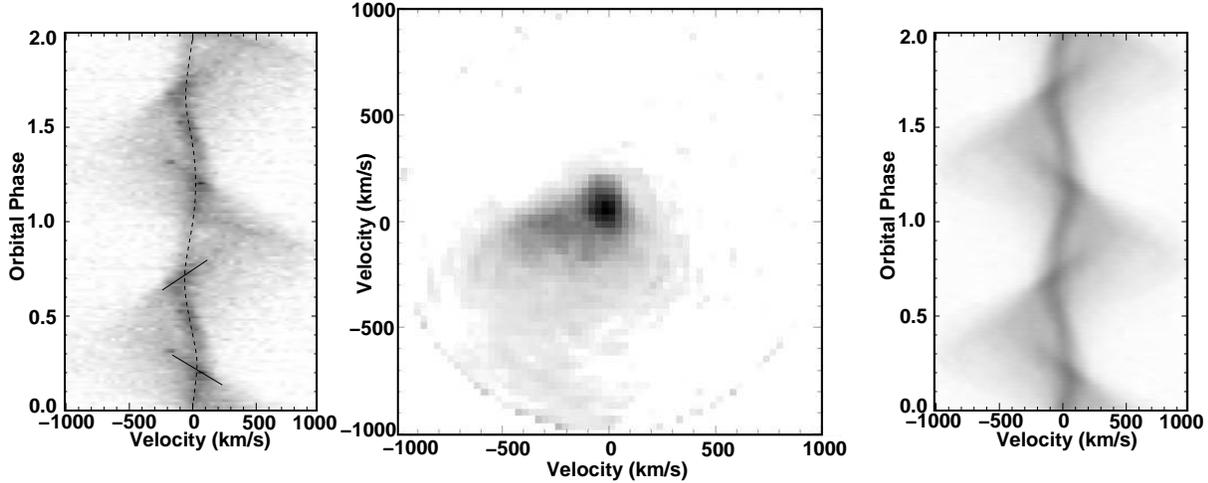} 
\caption{From left to right: HeII trailed spectra phase-folded on the
    derived ephemeris, the resulting Doppler tomogram and the
    reconstructed trailed spectra. }
\label{trl_dop}
\end{figure*}


Fig.~\ref{trl_dop} shows our HeII trailed spectra, the corresponding
Doppler tomogram (Marsh \& Horne 1988) and the reconstructed trailed
spectra of IGRJ14536-5522. The spectra were phase-folded on the
ephemeris derived above, continuum subtracted and then the Doppler
tomography code of Spruit (1998) was used.

We have also calculated \hbeta and \hgamma Doppler tomograms (not
shown).  These were found to be very similar to the HeII Doppler
tomogram, but the trailed spectra do not show the narrower components
as clearly.  Therefore we present only the HeII observations and the
corresponding interpretation.

The HeII trailed spectra clearly display the multiple components that
have been seen to some extent in other polars (e.g. Schwope et
al. 1995 and Rosen et al. 1996). From Fig.~\ref{trl_dop}, a narrow
component (indicated by the dashed curve) is visible throughout the
whole orbit and is generally recognised as emission from the
irradiated face of the secondary star. The fact that it is visible
throughout the whole orbit suggests a relatively low inclination,
although the radial velocity amplitudes are consistent with
more moderate inclinations (see e.g. V834 Cen: Potter et al. 2004).

There is possibly a second narrow component that is most visible where
it crosses and merges (indicated by diagonal lines) with the narrow
component from the secondary at phases $\sim 0.25$ and $\sim
0.75$. There also appears to be a broad but fainter underlining
component that reaches maximum blue-shift velocities of $\sim -1000$
km/s at phase $\sim 0.5$ and it remains discernible with maximum
redshift just before phase $\sim 1.0$. This fits the general picture
of emission from a ballistic accretion stream which then accelerates
and becomes magnetically channeled before reaching the surface of the
white dwarf (producing the broad component).

The HeII Doppler tomogram in Fig.~\ref{trl_dop} shows the typical
features of a moderately inclined polar. In particular, emission is
seen at the expected location of the irradiated face of the secondary,
approximately centered on x-velocity $\sim 0$ km/s and y-velocity
$\sim 100$ km/s. The x-velocity appears to be slightly negative offset
from 0. This can be explained as being due to the leading edge of the
secondary being more illuminated by the hot accretion region on the
surface of the white dwarf, which also leads in orbital phase
(e.g. V834 Cen: Potter et al 2004), although we cannot rule out that
it may also be as a result of inaccuracies in the determination of the
gamma velocity and/or the phasing of inferior conjunction on the
secondary.  The tomogram also shows emission at the expected location
of the ballistic accretion stream, seen to start at roughly the
location of the secondary and curving away towards more negative x and
y-velocities. There is also a very faint underlying component seen
protruding from zero velocities to low negative velocities.  This has
also been seen in other polars (e.g. Hu Aqr: Schwope, Mantel \& Horne
1997) and is generally explained as emission from accreting material where it
starts to be threaded by the magnetic field of the white dwarf.

\section{Photo-polarimetric analysis }

\begin{figure*}
\epsfxsize=16cm
\epsffile{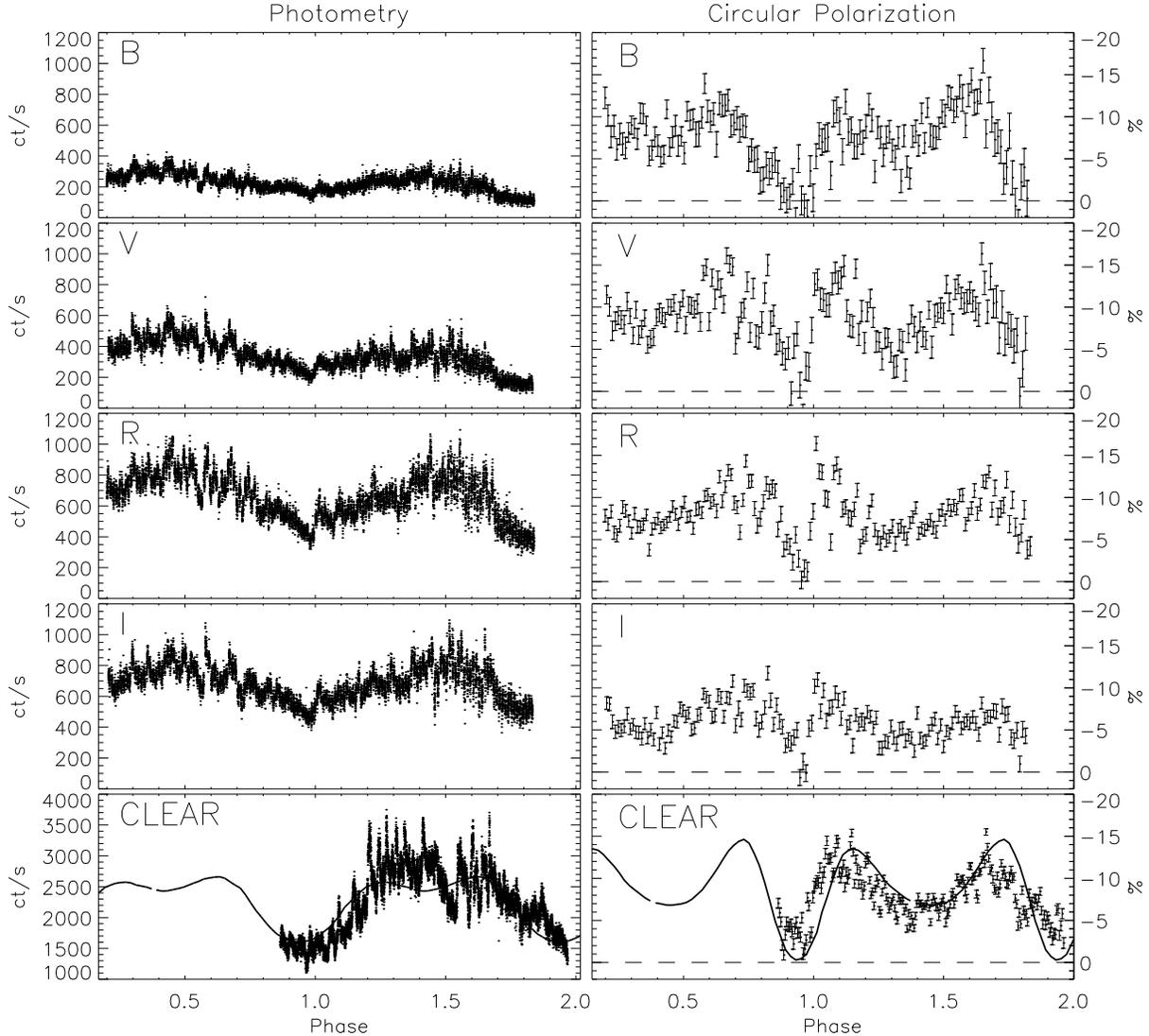} 
\caption{The photometric and polarimetric data phased on the
  spectroscopic ephemeris derived in section 4.2. The upper 4 plots
  are simultaneous B,V,R and I filtered observations of the 5th of
  April 2008 whilst the bottom plots were made with the clear filter
  on the 7th April 2008. Solid curves represent the model fit (see
  section 5.2). }
\label{pht_circ}
\end{figure*}

\subsection{The Photometry}

The left hand plots of Fig.~\ref{pht_circ} show the photometric
observations of IGRJ14536-5522 taken on the 5th (top 4 plots) and 7th
of April 2008 (last plot), phased on the spectroscopic ephemeris
derived in section 4.2. The 5th of April observations were obtained by
interleaving 60 second integrations between the B and V filters and
the R and I filters on channels 1 and 2 of the polarimeter
respectively. They are, therefore, effectively
simultaneous. Conditions were photometric during these
observations. The source was in a higher state.

\subsubsection{The orbital modulation and the dip}

As can be seen from Fig.~\ref{pht_circ}, the orbital modulation is
single-humped, which can be attributed to the beaming of cyclotron
emission from a single accretion spot on the surface of the white
dwarf. The orbital intensity minimum occurs when the emission region
is most face-on. The single humped morphology also implies a moderate
inclination for the system (see e.g. Ferrario \& Wickramasinghe (1990)
and Potter, Hakala \& Cropper (1998)).

The orbital intensity minimum is cut into by a narrow dip at
approximately phase $\sim 0.97$. This dip is also seen in the
polarimetry lightcurves (Fig.~\ref{pht_circ}, right). Considering the
accuracy of the phasing of the spectroscopic ephemeris (section 4.2),
the narrow dip could be due to an eclipse of the white dwarf by the
secondary star. However, this scenario is not consistent with the
results of our spectroscopic analysis and polarimetric modeling
(section 4.3 and 5.2 respectively), which favour moderate
inclinations. Therefore, it is more likely that the dip arises as a
result of absorption by material in the accretion stream and/or the
accretion column directly above the emission region when we see it
most face on (see e.g. Bridge et al. 2002).

\subsubsection{The photometric QPOs and flickering}

A close inspection of the photometry in Fig.~\ref{pht_circ} reveals
short period modulations throughout the orbit that are mostly
consistent with being due to noise or flickering. However, detailed
Fourier analysis of all of our photometry reveals that some of our
data sets show significant singularly persistent peaks that are
consistent with QPOs. 

An example data set is shown in the left (a) plot of
Fig.~\ref{28febQPO} from the 28th February 2008. We analysed the data
set by first splitting it into $\sim$40 minute sections with an
overlap of 85 percent. Next, each section of data was normalised to a
best fit second order polynomial before being subjected to Fourier
analysis. The results are shown as a trailed amplitude spectrum in the
left (b) plot of Fig.~\ref{28febQPO}. In this grey scale plot the
darker areas correspond to larger amplitudes. Between phases 0.2-1.0
the amplitude spectra do not show any significant peaks which
indicates that the variations are mostly flickering or noise. However,
there is a significant dominating signal centered on 0.0032(1) Hz (5.2
minutes, indicated by the dashed line) between phases 1.0 and
1.3. This is clear evidence of a QPO.  The 7 April 2008 data set also
shows clear evidence of a QPO (Fig.~\ref{7aprQPO}, left (a+b) plots)
at a similar period (0.0028(1) Hz, $\sim$5.9(3) minutes) centered on
the same phase range ($\sim$1.2) as well as a lower harmonic
(0.00136(15) Hz), both indicated with dashed lines.

In the left (c) plot of Fig.~\ref{28febQPO} we show the normalised
photometry during the phase range that is dominated by the
QPO. Overplotted are the least squares fit of the QPO frequencies. As
one can see, the 28 February photometric QPO is very well described by
the single dominant frequency as found in the trailed amplitude
spectra. The 7 April 2008 photometric QPO (Fig.~\ref{7aprQPO}, left
(c) plot) shows a more variable amplitude compared to the 28 February
QPO. This then explains why the trailed amplitude spectra shows two
frequencies for the QPO, one being the harmonic of the other. The left
(d) plots of Figs.~\ref{28febQPO} and \ref{7aprQPO} show the
corresponding amplitude spectra for the QPO dominated phase range.
The QPOs are discussed further in section 6.2.

\subsection{The Polarimetry}

The right hand plots of Fig.~\ref{pht_circ} show the polarimetric
(circular) observations of IGRJ14536-5522 taken on the 5th (top 4
plots) and 7th of April 2008 (bottom plot), phased on the
spectroscopic ephemeris derived in section 4.2. These were generated
from the same dataset as the photometry presented in the left hand
plots. We were unable to detect any linear polarization.

The circular polarization curve is double-humped and negative
throughout the entire orbit, which is consistent with emission from a
single accretion region in a moderately inclined system.  The
polarization reaches maximum negative values of $\sim$ 15 percent,
with the peaks of the humps occurring at orbital phases $\sim$ 0.1 and
$\sim$ 0.7.  There is also a narrow dip at phase $\sim$ 0.97, where
the polarization reaches a minimum of zero. This is coincident with
the narrow dip seen in the photometry, thus supporting the hypothesis
that the emission being absorbed is cyclotron radiation from the
accretion region (section 5.1.1). The phasing of the dip implies that
the location of the accretion region must be close to the line of
centers of the two stars. In addition, there is a broad minimum
centered on the narrow dip, which is typically caused by the beaming
of the cyclotron radiation at those phases where the line of sight
most closely approaches the axis of the column (Barret \& Chanmugam
1984; Wickramasinghe \& Meggitt 1985).

\subsubsection{The system geometry}

We investigated the system geometry further by modeling the
photo-polarimetry. We constructed a single arc model assuming
accretion along dipole field lines as in Potter et al (1997). A
magnetic field of 20MG was assumed and the cyclotron flux was
calculated using the stratified accretion shock grids of Potter,
Ramsay, Wu \& Cropper (2002). We considered only the clear filter data
because there does not appear to be any obvious wavelength dependence
in the multi-filtered observations. Consequently, different values for
the magnetic field strength and different shock models cannot be
investigated. However, the results of this model are largely
independent of our choice of the magnetic field strength.

The values for the parameters describing the inclination, dipole
offset angle and location, shape and size of the accretion region were
explored by comparing each resulting lightcurve to the data. The most
successful set of parameters integrated the emission from an arc
shaped region extending from 170$^{\rm o}$ to 230$^{\rm o}$ in
magnetic longitude and 10$^{\rm o}$ from the magnetic pole with a
system inclination of 50$^{\rm o}$. The magnetic dipole was offset by
an angle of 10$^{\rm o}$ from the white dwarf spin axis. A constant
unpolarized background of 1000ct/s was assumed.

We found that the model gave poor reproductions of the data for
inclinations outside a range of 45-55$^{\rm o}$, when either the
extent in phase or the double humped morphology of the circular
polarization were poorly reproduced. The solid curves in the lower 2
plots of Fig.~\ref{pht_circ} show the results of the best model
parameters over-plotted on the observations. The model has the
accretion region most face on at phase $\sim $ 0.95 and least face on
at phase $\sim $ 0.45.



\subsubsection{The polarized QPOs and flickering}

\begin{figure*}
\epsfxsize=18.5cm \epsffile{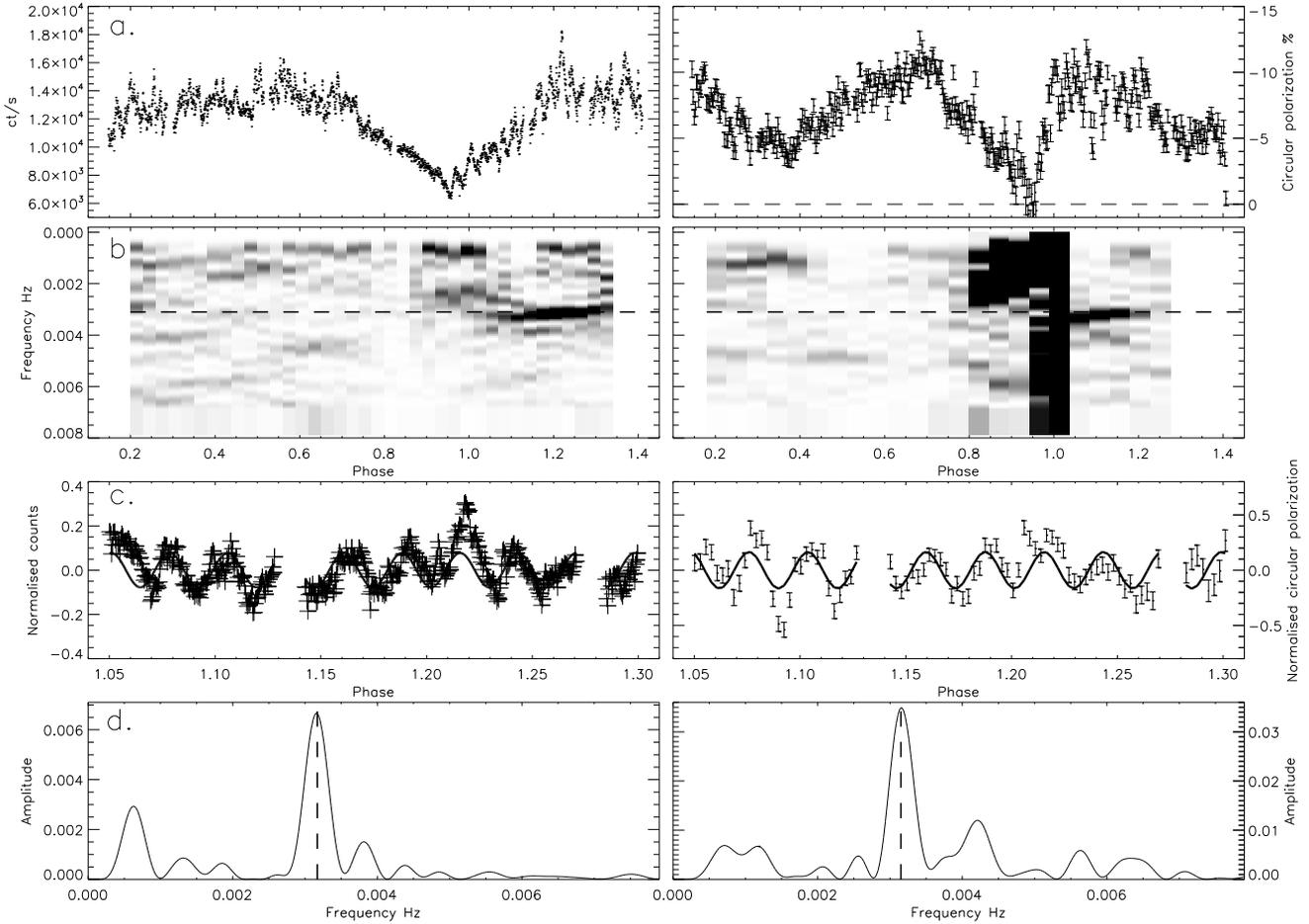}
\caption{Left plots, a-d: The photometry, the corresponding trailed
  amplitude spectra, the normalised photometry for the phase range
  1.0-1.3 and its corresponding amplitude spectra respectively. The
  solid curve is the least squares fit using the frequency derived
  from the trailed spectra (dashed line). Right plots: as in the left
  plots but for the circular polarization. Data set from 28 February
  2008.}
\label{28febQPO}
\end{figure*}

We analysed our circular polarimetric data in the same manner as that
of the photometry. The data and results for the 28 February 2008 are
shown in the right hand plots of Fig.~\ref{28febQPO}. As can be seen
from the right (b) plot, the trailed amplitude spectra is consistent
with mostly noise and/or flickering between phases 0.2-1.0. However a
QPO centered on 0.0031(1) Hz (5.4(3) minutes) is clearly evident
between phases 1.05-1.2. The QPO period is consistent within the
errors to that of the photometric QPO.

\begin{figure*}
\epsfxsize=18.5cm \epsffile{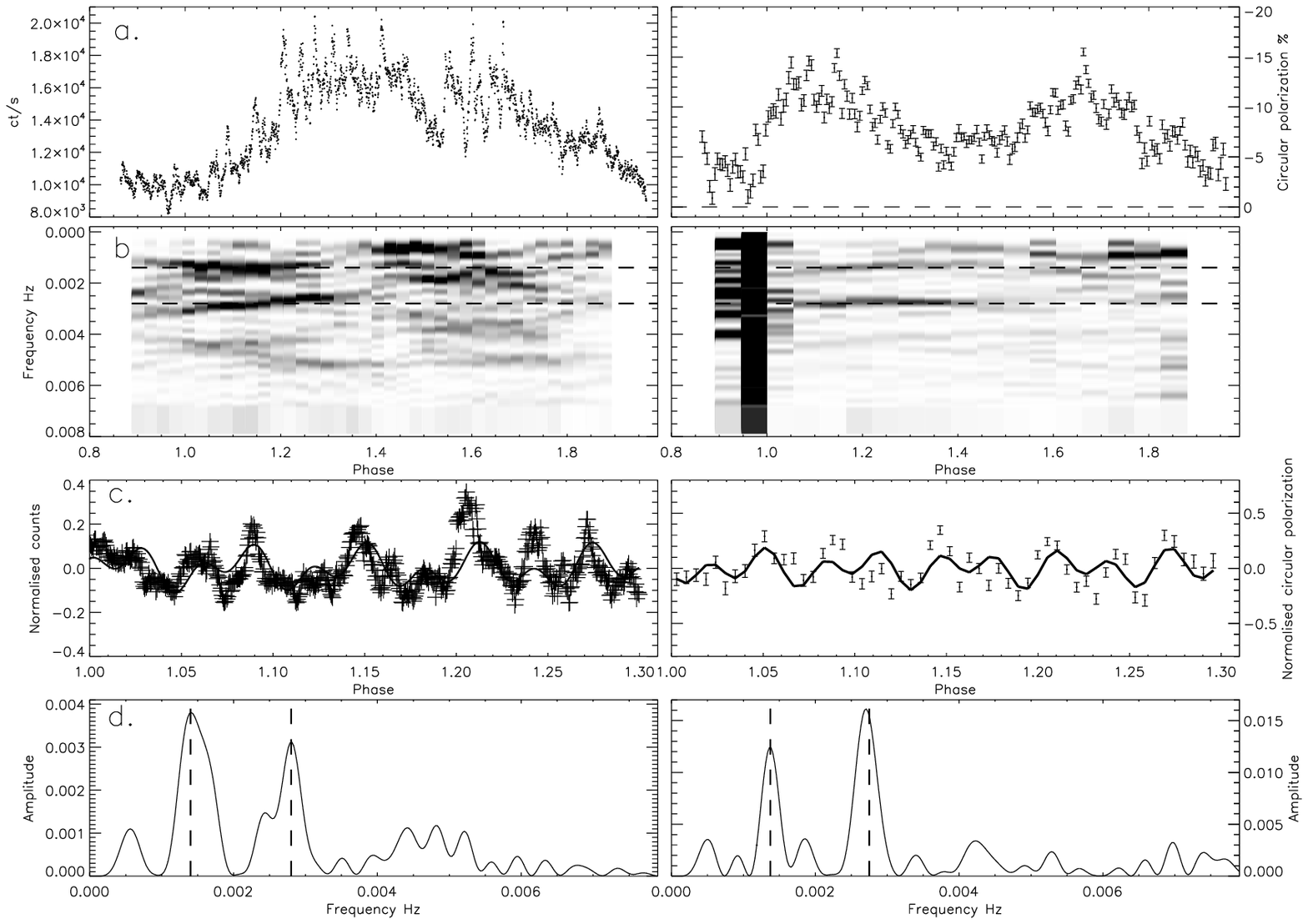}
\caption{As in Fig~\ref{28febQPO} for the 7 April 2008 data set.}
\label{7aprQPO}
\end{figure*}

The right hand plots of Fig.~\ref{7aprQPO} show the Fourier analysis
of the circular polarimetric data taken on 7 April 2008. Once again,
the trailed amplitude spectra displays the same characteristics as the
photometry, i.e. a dominant frequency is seen to be centered on
0.0028(1) Hz (5.9(3) minutes) and a lower harmonic at 0.0015(1) Hz
(both indicated by dashed lines) during the phase range 1.0-1.3.

In the right (c) plots of Fig.~\ref{28febQPO} and Fig.~\ref{7aprQPO}
we show the normalised circular polarimetry during the phase range
that is dominated by the QPOs. Overplotted are the least squares fit
of the QPO frequencies. As one can see, the circularly polarized QPOs
share the same characteristics as the photometric QPOs, namely the 28
February circularly polarized QPO is very well described by the single
dominant frequency found in the trailed amplitude spectra. Furthermore
the 7 April 2008 circularly polarized QPO shows a more variable
amplitude compared to the 28 February QPO, thus requiring a harmonic
frequency to better characterise the data. The QPOs are discussed
further in section 6.2.




\section{Discussion and Summary}

\subsection{Object classification and system geometry}

Our optical spectroscopy and high speed photo-polarimetry of the
INTEGRAL source IGRJ14536-5522 (=Swift J453.4-5524) unambiguously
confirm its identification as a polar. Negative circular polarization
is seen over all of the orbit, which is consistent with a single pole
accretor at a moderate inclination. We estimate some of the system
parameters by modeling the polarimetric observations. The most
successful model integrated the emission from an arc-shaped region
extending from 170$^{\rm o}$ to 230$^{\rm o}$ in magnetic longitude
and 10$^{\rm o}$ from the magnetic pole. The system inclination was
found to be in the range of 45-55$^{\rm o}$ with a magnetic dipole
offset angle of 10$^{\rm o}$.

\subsection{The QPOs and flickering}

Our high speed photo-polarimetry shows flickering on minute
time-scales. Furthermore, for the first time, we detect QPOs in the
photometry and circular polarimetry.



Photometric variations on the time scales of minutes are a common
feature of many polars, from X-rays to infrared (e.g. Szkody \& Margon
1980 and Watson, King \& Williams 1987), and have been characterised
as flickering, fluctuations, QPOs or erratic QPOs according to
different authors. In general (although not always strictly true),
QPOs describe variations that show some coherence over a period of
time. They mainly cluster in time-scales in the range of a few seconds
(1-5) or minutes (4-10). The (1-5s) QPOs are of low amplitude and have
been observed in the optical in only a few polars (e.g. V834 Cen, AN
Uma and VV Pup). The larger amplitude (4-10 min) QPOs seem to be a
general feature of most polars. By eye, they can appear to show some
sort of coherence, but fail to show any significant singularly
persistent peaks under Fourier analysis. Instead, groupings of periods
are seen and are often referred to as QPO-like. Flickering and
fluctuations best describe variations that do not show any periodic
behavior.

There is evidence that the 4-10 minute QPOs may consist of the
superposition of regular periodic oscillations: Bonnet-Bidaud, Somova
\& Somov (1991) report on observations of AM Her during an
intermediate brightness state where they find fluctuations to be
nearly periodic, with a period increasing from 250s to 280s. They
reason that they observed AM Her in a transition from bright-state to
faint-state where a given accretion rate is able to excite stable
oscillations. Alternatively, during the higher state, many
oscillations may be present but masked by the superposition of
different modes corresponding to different accretion tubes. Therefore,
during the transition, the accretion rate decreases, reducing the
number of accretion tubes until finally only one unique tube
contributes to the emission.

Ultimately, the photometric oscillations are caused by variations in
the accretion flow. There are several theories/models that attempt to
explain the variations. King (1989) remarks that the (4-10 minute)
periods are characteristic of a dynamical time at the photosphere of
the companion. Irradiation of the region near the L$_{1}$ point will
result in the formation of an ionisation front, which will tend to
oscillate and therefore modulate the accretion. Other models involve
the capture of inhomogeneities of the accretion flow at the capture
radius (Kuijpers and Pringle 1982) and accretion gate mechanisms
(Patterson et al. 1981).

The short period QPOs (1-5s) have been observed in $\sim $ 6 systems
(e.g. V834 Cen, AN Uma and VV Pup).  Langer, Chanmugam \& Shaviv
1982) realised that the 1-5s oscillations were consistent with the
cooling time-scales for white dwarf-radiated shock waves and thus QPOs
are potentially powerful probes of the radiative shocks. Observations
of VV Pup (Larson 1989) demonstrated conclusively that the QPOs arose
in a region near the shock emission region on the white dwarf.


From our results presented in Figs.~\ref{28febQPO} and \ref{7aprQPO}
it is clear that IGRJ14536-5522 exhibited photometric and polarized
QPOs (5-6 minutes) on two separate occasions. Additionally, the fact
that we observe the QPOs oscillations in the polarimetry,
unequivocally places the emission site at the cyclotron emitting shock
region. Furthermore, the QPOs were seen during phases 1.0-1.3 only. We
propose that the super-position of the oscillations arising from
numerous accretion tubes (similar to the mechanism of Bonnet-Bidaud,
Somova \& Somov 1991) applies to IGRJ14536-5522. However, instead of a
reduced accretion state leading to a reduced number of accretion
tubes, we suggest that for most of the orbit the superposition of
emission from many accretion tubes leads to the observed
flickering. However during the phase interval 1.0-1.3 we
preferentially see the trailing edge of the accretion region as the
white dwarf rotates. The leading part of the accretion region is
effectively shielded from our view by its trailing edge for this phase
interval. Consequently we observe a singular QPO corresponding to a
singular accretion tube.

Our observations of IGRJ14536-5522 indicate that it undergoes
relatively frequent changes in accretion state, making it a good
candidate for capturing it in transition and testing this scenario
further. If QPOs are also present in the X-rays then IGRJ14536-5522 is
an ideal source for investigating the two dominating shock cooling
mechanisms: i.e. bremsstrahlung and cyclotron cooling.

Finally, we would like to note the importance of optical follow-up
observations of candidate CV INTEGRAL sources, in particular with
photometry and/or polarimetry. It has been noted that many CVs
detected by INTEGRAL are IPs, both new and re-discoveries (see
e.g. Barlow et al. 2006 and Revnivtsev et al 2008). However, as
Pretorius (2009) points out, it is only through follow-up observations
that unequivocal identifications can be made. A case in point is
IGRJ14536-5522, which had been assumed to be an IP (see
e.g. Revnivtsev et al. 2008) even though a spin period had not been
detected.






\section{Acknowledgments}

We thank the anonymous referee for comments and suggestions that have
significantly improved the paper.

This material is based upon work supported financially by the National
Research Foundation.  Any opinions, findings and conclusions or
recommendations expressed in this material are those of the author(s)
and therefore the NRF does not accept any liability in regard to
thereto.

Some of the observations reported in this paper were obtained with the
Southern African Large Telescope (SALT), a consortium consisting of
the National Research Foundation of South Africa, Nicholas Copernicus
Astronomical Center of the Polish Academy of Sciences, Hobby Eberly
Telescope Founding Institutions, Rutgers University,
Georg-August-Universität Göttingen, University of Wisconsin - Madison,
Carnegie Mellon University, University of Canterbury, United Kingdom
SALT Consortium, University of North Carolina - Chapel Hill, Dartmouth
College, American Museum of Natural History and the Inter-University
Centre for Astronomy and Astrophysics, India.

\end{document}